\documentclass{article}

\usepackage{fullpage}
\usepackage[english]{babel}
\usepackage{verbatim}

\usepackage{amsmath}
\usepackage{amssymb}
\usepackage{amsthm}
\usepackage{algorithm}
\usepackage{algorithmic}
\usepackage{graphicx}
\newtheorem{theorem}{Theorem}

\title{Trembling hand perfection is NP-hard\thanks{Work supported by {\em Center for
      Algorithmic Game Theory}, funded by the Carlsberg
    Foundation.} \\ (note)}
\author{Peter Bro Miltersen}
\begin{document}
\maketitle
\begin{abstract}
It is NP-hard to decide if a given pure strategy Nash equilibrium of
a given three-player game in strategic form with integer payoffs
is trembling hand perfect.
\end{abstract}
\section{Introduction} 
Trembling hand perfection \cite{Selten} is a well-established
refinement of Nash equilibrium.
We prove:
\begin{theorem}
It is NP-hard to decide if a given pure strategy Nash equilibrium of
a given three-player game in strategic form is trembling hand perfect.
\end{theorem}
In particular, unless P=NP, there is no polynomial time algorithm
for deciding if a given equilibrium of a given three-player game
in strategic form is trembling hand perfect. 
Arguably, this can be interpreted as a deficiency of the trembling
hand solution concept. 

Note that in contrast to the above hardness result, 
one may efficiently determine if a given equilibrium
of a {\em two}-player game is trembling hand perfect. Indeed,
for the two-player case, an equilibrium is trembling hand perfect
if and only if it is undominated. This can be checked
by linear programming in polynomial time.

The proof below can be rather easily modified to show that 
it is NP-hard to decide if a
given equilibrium of a three-player game is {\em proper}
\cite{Myerson}. 
For properness,
we do not know if the two-player case is easy or not.

Finally, we remark that we do not know if it is {\em in} NP to
decide if a given equilibrium is perfect (or proper). It seems
that an obvious
nondeterministic algorithm would be to guess a 
{\em lexicographic belief structure}
and appeal to the characterizations of Blume {\em et al} \cite{Blume}
and Govindan and Klumpp \cite{Govindan} of trembling hand perfection
in terms of these. However, we do not know if 
a lexicographic belief structure witnessing perfection (or properness)
can be represented as a polynomial length string over a finite alphabet.

\section{Proof}
Our proof is a reduction from the problem of approximately computing
minmax values of 3-player games with 0-1 payoffs, 
a problem that was recently shown
to be NP-hard by Borgs {\em et al} \cite{Borgs}. In particular, it follows
from Borgs {\em et al.} that the following {\em promise problem}
MINMAX is NP-hard:

\vspace*{0.2cm}

\noindent
MINMAX: 
\begin{enumerate}
\item{}YES-instances:
 Pairs $(G,r)$ for which the minmax value for Player 1 in the 3-player
game $G$ is strictly smaller than the rational number $r$.
\item{}NO-instances:
 Pairs $(G,r)$ for which the minmax value for Player 1 in $G$ 
is strictly greater than $r$.
\end{enumerate}
In fact, by multiplying the payoffs of the game with the denominator
of $r$, we can without loss of generality assume that $r$ is an integer.
We now reduce MINMAX to deciding trembling hand perfection.

Let $G$ be a three-player game in strategic form and let $r$ be an
integer.
We define $G'$ be the game where the strategy space of each player
is as in $G$, except that it is extended by a single pure strategy, $\bot$.
The payoffs of $G'$ are defined as follow. The payoff
to Players 2 and 3 are 0 for all strategy combinations.
The payoff to Player 1 is $r$ for all strategy combinations where
at least one player plays $\bot$. For those strategy combinations
where no player plays $\bot$, the payoff to player 1 is the same
as it would have been in the game $G$. Obviously, $\mu = (\bot, \bot, \bot)$
is a Nash equilibrium of $G'$.

We claim that if the minmax value for Player 1 in $G$ is strictly 
smaller than $r$, then
$\mu$ is a trembling hand perfect equilibrium of
$G'$. 
Indeed, let $(\tau_2, \tau_3)$ be a minmax strategy
profile of Players 2 and 3 in $G$. Let $\tau$ be any profile of
$G'$ where Players 2 and 3 play $(\tau_2, \tau_3)$.
Also, let $u$ be the strategy profile of $G'$ where
each player mixes all pure strategies uniformly. Now define
\[ \sigma_k = (1-\frac{1}{k}-\frac{1}{k^2})\mu + \frac{1}{k} \tau + \frac{1}{k^2} u \]
We have that $\sigma_k$ is a fully mixed strategy profile of $G'$ 
converging to $\mu$ as
$k \rightarrow \infty$. Also, for sufficiently large $k$, the strategies
of $\mu$ are best replies to $\sigma_k$. This follows from the fact
that Players 2 and 3 are indifferent about the outcome and the
fact that Player 1 gets payoff $r$ by playing $\bot$ while he gets
a payoff strictly smaller than $r$ for large values of $k$ by playing
any other strategy.
We conclude that $\mu$ is trembling hand perfect, as desired.

On the other hand, we claim that
if the minmax value for Player 1 in $G$ is strictly greater than $r$, then
$\mu$ is a {\em not} a trembling hand perfect equilibrium of
$G'$. 
Indeed, let 
$(\sigma_{k,1}, \sigma_{k,2}, \sigma_{k,3})_k$ be any sequence of fully
mixed strategy profiles converging to $(\bot, \bot, \bot)$.
Since $\sigma_{k,2}$ and $\sigma_{k,3}$ do not put all their probability
mass on $\bot$, Player 1 has a reply to $(\sigma_{k,2}, \sigma_{k,3})$
with an expected payoff strictly greater than $r$. Therefore,
$\bot$ is not a best reply of Player 1 to $(\sigma_{k,2}, \sigma_{k,3})$ and
we conclude that $(\bot, \bot, \bot)$ is not trembling hand
perfect.

That is, we have reduced the promise problem MINMAX to deciding trembling
hand perfection and are done.
\bibliographystyle{plain}

\end{document}